\documentclass[10pt,conference]{IEEEtran}
\usepackage{cite}
\usepackage{mathrsfs}
\usepackage{color,soul}
\usepackage{graphicx}
\usepackage{epsfig}
\usepackage[usestackEOL]{stackengine}
\usepackage{flushend}



\makeatletter
\newcommand*\titleheader[1]{\gdef\@titleheader{#1}}
\AtBeginDocument{%
	\let\st@red@title\@title%
	\def\@title{%
		\bgroup\normalfont\large\centering\@titleheader\par\egroup
		\vskip2em\st@red@title}
}
\makeatother

\title{Help Me Find a Job: A Graph-based Approach for Job Recommendation at Scale}
\titleheader{\vspace{-0ex}Accepted at 2017 IEEE International Conference on Big Data (BIGDATA)}

\begin{document}

\DeclareRobustCommand*{\IEEEauthorrefmark}[1]{%
	\raisebox{0pt}[0pt][0pt]{\textsuperscript{\footnotesize #1}}%
}

\author{
	\IEEEauthorblockN{Walid Shalaby\IEEEauthorrefmark{1}, BahaaEddin AlAila\IEEEauthorrefmark{2}, Mohammed Korayem\IEEEauthorrefmark{3},\\ Layla Pournajaf\IEEEauthorrefmark{3}, Khalifeh AlJadda\IEEEauthorrefmark{3}, Shannon Quinn\IEEEauthorrefmark{2}, and Wlodek Zadrozny\IEEEauthorrefmark{1}}\\ 
	\IEEEauthorblockA{\IEEEauthorrefmark{1}Department of Computer Science, University of North Carolina at Charlotte \\
	\small \{wshalaby, wzadrozn\}@uncc.edu \\\\
	\IEEEauthorrefmark{2}Institute for Artificial Intelligence, University of Georgia\\ 
	\small \{bahaaeddin.alaila, squinn\}@uga.edu \\\\
	\IEEEauthorrefmark{3}Search Data Science, CareerBuilder \\
	\small \{mohammed.korayem, layla.pournajaf, khalifeh.aljadda\}@careerbuilder.com} 	
}

%\author{
%	\IEEEauthorblockN{Walid Shalaby\IEEEauthorrefmark{1},
%		BahaaEddin AlAila\IEEEauthorrefmark{2},
%		Mohammed Korayem\IEEEauthorrefmark{3}, 
%			Layla Pournajaf\IEEEauthorrefmark{3},
%			Khalifeh AlJadda\IEEEauthorrefmark{3}, and
%	 Wlodek Zadrozny\IEEEauthorrefmark{1}\\IEEEauthorrefmark{1}}
%	\IEEEauthorblockA{\IEEEauthorrefmark{1}Department of Computer Science, UNC Charlotte,
%		Charlotte, NC, USA\\\IEEEauthorrefmark{2}Institute for Artificial Intelligence, University of Georgia  \\\IEEEauthorrefmark{3}CareerBuilder, %\\
%		Norcross, GA, USA\\ wshalaby,wzadrozn@uncc.edu, bahaaeddin.alaila@uga.edu,\\mohammed.korayem, ayla.pournaja,khalifeh.aljadda@careerbuilder.com}

% There's nothing stopping you putting the seventh, eighth, etc.
% author on the opening page (as the 'third row') but we ask,
% for aesthetic reasons that you place these 'additional authors'
% in the \additional authors block, viz.
% Just remember to make sure that the TOTAL number of authors
% is the number that will appear on the first page PLUS the
% number that will appear in the \additionalauthors section.

\IEEEoverridecommandlockouts
\IEEEpubid{\makebox[\columnwidth]{978-1-5386-2715-0/17/\$31.00~
		\copyright2017
		IEEE \hfill} \hspace{\columnsep}\makebox[\columnwidth]{ }}

\maketitle
\begin{abstract}
Online job boards are one of the central components of modern recruitment industry. With millions of candidates browsing through job postings everyday, the need for accurate, effective, meaningful, and transparent job recommendations is apparent more than ever. While recommendation systems are successfully advancing in variety of online domains by creating social and commercial value, the job recommendation domain is less explored. Existing systems are mostly focused on content analysis of resumes and job descriptions, relying heavily on the accuracy and coverage of the semantic analysis and modeling of the content in which case, they end up usually suffering from rigidity and the lack of implicit semantic relations that are uncovered from users' behavior and could be captured by Collaborative Filtering (CF) methods. Few works which utilize CF do not address the scalability challenges of real-world systems and the problem of cold-start. In this paper, we propose a scalable item-based recommendation system for online job recommendations. Our approach overcomes the major challenges of sparsity and scalability by leveraging a directed graph of jobs connected by multi-edges representing various behavioral and contextual similarity signals. The short lived nature of the items (jobs) in the system and the rapid rate in which new users and jobs enter the system make the cold-start a serious problem hindering CF methods. We address this problem by harnessing the power of deep learning in addition to user behavior to serve hybrid recommendations. Our technique has been leveraged by CareerBuilder.com which is one of the largest job boards in the world to generate high-quality recommendations for millions of users. 

\end{abstract}

\section{Introduction}
Recommendation systems have demonstrated success in many online domains such as sales, media, and social communities by connecting users to the items of their interest and building their loyalty. With the excess of available online information, job seekers need to have access to relevant job openings in almost real-time, however, browsing through thousands of jobs for finding few relevant ones can be a tedious task for many applicants. With this motivation, our goal is to build an effective recommendation system to improve the job search process by harnessing multiple signals of relevance and providing job seekers (i.e. applicants) with personalized job recommendations. 
% 
%In CareerBuilder, we have access to blah blah
%building job seekers loyalty
%
%converting one-time users to registered users
Collaborative filtering (CF) is one of the widely used recommendation approaches which exploits the user-item interactions to identify similar items (a.k.a. item-based CF) or users (a.k.a. user-based CF) and predict a user's future interests~\cite{herlocker2004evaluating}. However, there are inherent challenges in building such systems in the domain of job search and recommendations. 
%\begin{itemize}
% Scalability and Incremental updates
%Active jobs vs expired jobs
\paragraph{Scalability} Building a scalable recommendation system for millions of users and jobs is crucial. 
%(it would show the gap) Existing job recommendation systems belong to either content-based applicant-job matcher approaches~\cite{guo2016resumatcher,diaby2013toward} or user-based methods~\cite{rafter2000automated,lu2013recommender}. 
In the recent years, item-based recommendation systems have gained more popularity as they are more scalable compared to their user-based counterparts~\cite{deshpande2004item,sarwar2001item}. However, with the vast amount of incoming jobs everyday, building and maintaining a job-based system is not trivial. We propose a graph-based structure in order to efficiently model job-job relationships with variable-length neighborhood sizes. For further performance boost through incremental updates, we utilize active jobs in building and updating the online graph while leveraging expired jobs offline.
 % Explicit vs implicit data
% Sparsity
\paragraph{Job Similarities/Sparsity} User behaviors are typically captured by user-item interactions expressed as ratings in domains such as movie recommendations. These ratings are then used in computing a variety of similarity metrics between items or users; namely cosine or correlation-based measures~\cite{adomavicius2005toward}. However, asking users to rate or rank jobs based on their relevance is not realistic or applicable in a real-world recruiting systems. Online recommendation systems use variety of explicit and implicit information sources such as purchasing histories in e-commerce systems~\cite{li2009grocery,linden2003amazon}, browsing and clicking behaviors in news recommendations~\cite{das2007google}, or views in online video recommendation~\cite{baluja2008video}. One major drawback of relying on such data sources alone is the high level of data sparsity in which typical item-item similarity measures may fail~\cite{yildirim2008random}. Hence, we explore various data sources in the jobs domain and develop alternative similarity measures to fully capture the relationships between jobs, and alleviate the data sparsity problem. %Furthermore, the role of other implicit data should be also investigated. %Recently, graph-based models try to alleviate this problem using random walks or other preference propagation methods~\cite{haveliwala2002topic,yildirim2008random,baluja2008video}. 
%'cold start' problem.
\paragraph{Cold-Start} Relying on behavioral data solely results in lower quality recommendations when there is no user-job interaction data. This is akin to the 'cold-start' problem which becomes more prominent in dynamic systems such as job recommendations where new users and/or jobs are introduced to the system at high rates.
%\textcolor{blue}{Yet, it doesn't include millions of unregistered users who interact with jobs sporadically\footnote{i don't get your point here}}. 
Moreover, many users become inactive when they do not interact with the jobs for a considerable period of time. To get the attention of such users and get them to reengage with the system and become active users again, it is necessary to distinguish between them and brand new users. 
%Content rich domains such as news~\cite{claypool1999combining} benefit from hybrid recommendation systems which incorporate content-based methods to tackle the cold start problem~\cite{adomavicius2005toward}.  

%\end{itemize}
%\paragraph{***Kind of a summary with highlighted contributions}  
%Some details should be removed/adde
In this paper, we propose a novel item-based job recommendation system to overcome the above challenges. Our model aims at filling the gap in existing job recommendation systems by leveraging multiple contextual and behavioral signals in one unified salable graph-based architecture. At a high level, our system consists of three main steps. First, we build a scalable homogeneous graph of jobs as nodes with multiple association edges between nodes capturing multiple behavioral and contextual signals. Second, a weighted directed edge is created for every job pair through weighted aggregation of all their association edges. Third, we model user preferences and generate recommendations using propagation-based search strategies on the graph. Our main contributions can be summarized as below:
\begin{itemize}
%\noindent
\item We propose various item-item association measures tailored for job recommendations including: 
\begin{itemize}
\item several symmetric and asymmetric scores based on collected clicks and applications data (from CareerBuilder.com) to capture the implicit and explicit user behaviors without bias towards certain types of users or jobs.  
\item a content-based similarity measure which is learned by creating neural job embeddings using job descriptions~\cite{yuan2016solving}.
\end{itemize}
%\noindent
\item We successfully model user preferences in terms of active jobs or predefined job categories. By recognizing the differences between \textit{active} vs \textit{passive} users as well as \textit{registered} vs \textit{unregistered} users, we are able to provide some level of personalization for a variety of 'cold-start' users with limited or no history. %based on the availability of data, we build user's preference vector in the terms of active jobs or approximate it in the terms of job categories.
%\noindent
\item Our model utilizes a local propagation-based search strategy to generate personalized recommendations even when data is scarce. To further alleviate the 'cold-start' users, we also adopt a separate personalized strategy based on a modified PageRank algorithm %within clusters of jobs 
to improve our recommendations by incorporating the popularity of jobs.     
%\noindent
\item Experimental results show that our approach outperforms the classical CF in terms of recommendations quality and relevancy. In addition, our proposed architecture is generic and can be easily applied to recommendation domains other than recruitment services.
%\item We  during modeling the user preferences and  
%
%Our proposed method ranks the relevant jobs for the user by taking into account the user activity with the expired jobs in the past and/or user skills (i.e. the resume content).
%
%***This part provides too much details in the introduction?
%\begin{itemize}
%\item First, we classify the current pool of jobs into pre-defined job categories~\cite{}. We also classify users into job categories exploiting their past activity over expired jobs. 
%We map the expired jobs to the most similar jobs in the current pool using a content-based job matching algorithm and use their category as the user category. In the case of brand new users with zero activity, we classify them based on the content of their resume. 
%\item Then, we use the user category to extract a personalized subgraph of current graph of jobs. 
%\item Finally, we apply the PageRank algorithm on the generated subgraph to rank the relevant jobs for the user. 
%By shrinking the size of the nodes in the graph, we avoid the bias towards the more popular jobs and achieve the desired personalization for the user. 
%\end{itemize}
\end{itemize}

The remaining sections are organized as follows: Section 2
reviews the related work in detail. In Section 3, we introduce our proposed job recommendation system. Section 4 presents the implementation details of our system for Careerbuilder job recommendations, and the evaluation results. Finally,
we discuss conclusions and future work in Section 5.

%\textcolor{red}{
%**** (Maybe Move this to Systems regarding to Mohammad's note in red)
%Although, we can still approximate users preference vectors for inactive/passive users using their old history with expired jobs or their resume contents, such vector would not fully capture their real-time preferences. Hence, we extract a semi-personalized subgraph of current graph using the top-k categories of the user's approximate preference vector. Then, we apply PageRank on this subgraph to further improve our recommendations by incorporating the general popularity of jobs within each category. }

\section{Related Work}

Collaborative filtering (CF) methods rely solely on the past activity of users (e.g. ratings, purchases) for generating recommendations. Existing CF methods belong to two categories: i) \textit{memory-based} (a.k.a. heuristic-based), and ii) \textit{model-based} methods~\cite{breese1998empirical,adomavicius2005toward}. For any unseen user-item pairs, memory-based methods utilize a heuristic-based search strategy to find users with similar taste to the current user. Then, they predict the user's action on the new item using the aggregated behavior of these similar users. Alternatively, in an item-based variation, the past behavior of the user on similar items is utilized. On the other hand, model-based methods focus on learning an off-line model from the past activity of users using machine learning techniques such as clustering~\cite{o1999clustering,shepitsen2008personalized} or matrix factorization~\cite{koren2009matrix}. A comprehensive survey of CF-based methods can be found in~\cite{herlocker2004evaluating,candillier2007comparing}.

CF is widely adopted by real-world systems since it is suitable for a variety of applications regardless of the underlying domains. More specifically, item-based CF has shown to be more scalable and results in higher quality recommendations compared to the user-based methods~\cite{deshpande2004item,sarwar2001item}; hence, it is more suitable for online systems with millions of users and fewer number of distinct items. One of the shortcomings associated with CF involves scenarios in which users or items do not have enough interaction data (e.g. users with unusual tastes, new users or items). One way to overcome this shortcoming is to exploit another category of recommendation systems known as~\textit{content-based methods} which build models using the explicit domain-specific features of users and items~\cite{lops2011content}. Systems that combine both CF and content-based approaches are known as \textit{hybrid recommendation systems}. 

Since our proposed method incorporates content-based signals to the CF-based core system, it is considered a hybrid method. We utilize content in two different scenarios by distinguishing between the 'cold-start' situation for jobs and users. In the case of a new job, we compute a content-based similarity measure between jobs by building a neural network that learns job embeddings~\cite{yuan2016solving}. This way, we guarantee an association between jobs even when they do not share any user interactions. For users who do not have any activities, we model their preferences by classifying their resume content into pre-defined fine-grained job categories, and then recommend them popular jobs under these categories.

\subsection{Graph-based Recommendation Systems}
Graph-based models adopt link analysis methods from graph theory to address the shortcomings of CF-based approaches such as sparsity and improve the quality of the recommendations~\cite{chen2005link}. One of the main research directions in this area emerged after the success of the  PageRank method for ranking of web pages based on their importance and popularity~\cite{page1999pagerank}. PageRank builds a directed graph of webpages as nodes with edges representing the transition probability between nodes. By adopting a memory-less strategy, they capture the importance of each webpage based on the single hop incoming links to them. The intuition is that a random surfer already visiting a page would choose the next webpage, mostly an adjacent one, based on the transition probabilities between the pages, with a small chance of hopping to a random webpage. Graph-based recommendation systems are differentiated based on how they build the graph and traverse it for recommendations.~\textit{Heterogeneous} graph-based models build a bipartite graph of both users and items~\cite{zhou2007bipartite}, while~\textit{homogeneous} models only include users~\cite{aggarwal1999horting} or items~\cite{yildirim2008random} as nodes.

The CF approach can be expressed as a link prediction problem in a user-item bipartite graph in which edges reflect the interaction between users and items. To predict a non-existing edge between a user and an item, a variety of neighbor-based and path-based linkage measures are adapted from graph theory~\cite{chen2005link}. In a recent study by Li and Dias ~\cite{li2009grocery}, bipartite graphs are used for grocery product recommendations in which nodes include both products and consumers, while edges represent the frequencies of product purchases by consumers. The authors proposed an item similarity measure based on the aggregated higher order stochastic transition matrices between items to overcome the sparsity problem. Then, they generalized a random-walk-with-restart method~\cite{pan2004automatic} for personalized recommendations based on the content of the shopping basket. Our method differs from these approaches in that we only build an item-based graph for a more scalable solution. 
 
A homogeneous graph-based method is proposed by Aggarwal et al.~\cite{aggarwal1999horting}, with users as the nodes and a predictability measure between users as the edges. In contrast to nearest neighbor search, starting from the user, they traverse paths from each connected node to find a user that has rated the item of interest. Similar to other user-based CF approaches, this method suffers from scalability problem when the number of users becomes very large. Alternatively, some methods build a graph of items~\cite{gori2006research,baluja2008video}. Baluja et al.~\cite{baluja2008video} build a graph of Youtube videos (as items) with edges indicating the co-views of video pairs by the same users. They adopt a semi-supervised label propagation method to recommend videos to users. In our work, we also build an item-based graph, however, instead of a global propagation-based search on the graph, we mainly use a local search strategy. Moreover, we allow multiple edges between the nodes and compute an aggregated asymmetric association score based on job contents, applications (co-apps), and co-clicks to capture the popularity of jobs as well as their content and interaction similarities. 
 
Several methods has adapted PageRank for personalized recommendations~\cite{haveliwala2003analytical,kim2011personalized,gori2006research}. Topic-sensitive PageRank~\cite{haveliwala2002topic} uses a limited set of predefined topics and determines the probabilistic membership of each page into these topics. It modifies the PageRank random walk to consider hopping between topics as well as pages, then uses the classification of the current page to compute personalized weights for each topic. Gori and Pucci~\cite{gori2006research} proposed a research paper recommendation system which uses a modified PageRank algorithm to rank papers in a citation graph biased by user preferences. In this work, we adopt a personalized PageRank approach when data is less than plentiful to get the full advantages of our heuristic-based search method.  

\subsection{Job Recommendation Systems}
Information extraction from resumes and job descriptions is one of the main areas of research in jobs and recruitment industry~\cite{zhao2015skill,singh2010prospect,guo2016resumatcher}. These works involve text mining, skill normalization, and developing similarity metrics for matching jobs and candidate profiles. Most of the existing recommendation systems in this domain focus on candidate selection by human resources rather than attracting job seekers through job recommendations which is the main goal of our paper~\cite{lee2007architecture,farber2003automated}. Existing automated job recommendation systems belong to either content-based applicant-job matcher approaches~\cite{guo2016resumatcher,diaby2013toward}, or user-based methods~\cite{rafter2000automated,lu2013recommender}.

Rafter et al.~\cite{rafter2000automated} proposed a user-based CF system for job recommendation in the JobFinder website. the authors utilized the overlap of interacted jobs as the similarity measure between two users. Then, they apply a nearest neighbor search and a clustering-based approach to generate recommendations. Malinowski et al.~\cite{malinowski2006matching} proposed a probabilistic method to match jobs and resume profiles for producing both candidate and job recommendations. 
A recent work by Paparrizos et al.~\cite{paparrizos2011machine} predicts the next job transition of users by building a supervised learning model using their past employment history. This model recommends the predicted institution to the user and can not be considered a real-time job recommendation system. 

Several works exploit social networks to generate job recommendations~\cite{diaby2013toward,lu2013recommender}. Lu et al.~\cite{lu2013recommender} proposed a graph-based method with three types of nodes including users, companies, and jobs to generate job recommendations exploiting the relationships between all the three entities in a social network. The authors only present preliminary results based on a small sample of the data and further evaluations using online data is left as future work. In our work, we also propose a graph-based method exploiting user-job interactions and job postings content, however, in addition to accuracy, one of the main focuses of our work is scalability and real-time recommendations which is not commonly addressed in the previous work.
%Must describe implementation of a system that solves a significant real-world problem. The focus should be on describing the problem, its significance, decisions and tradeoffs made when making design choices for the solution, deployment challenges, and lessons learned.
\section{System Description  }
\begin{figure}
	\centering \epsfig{file=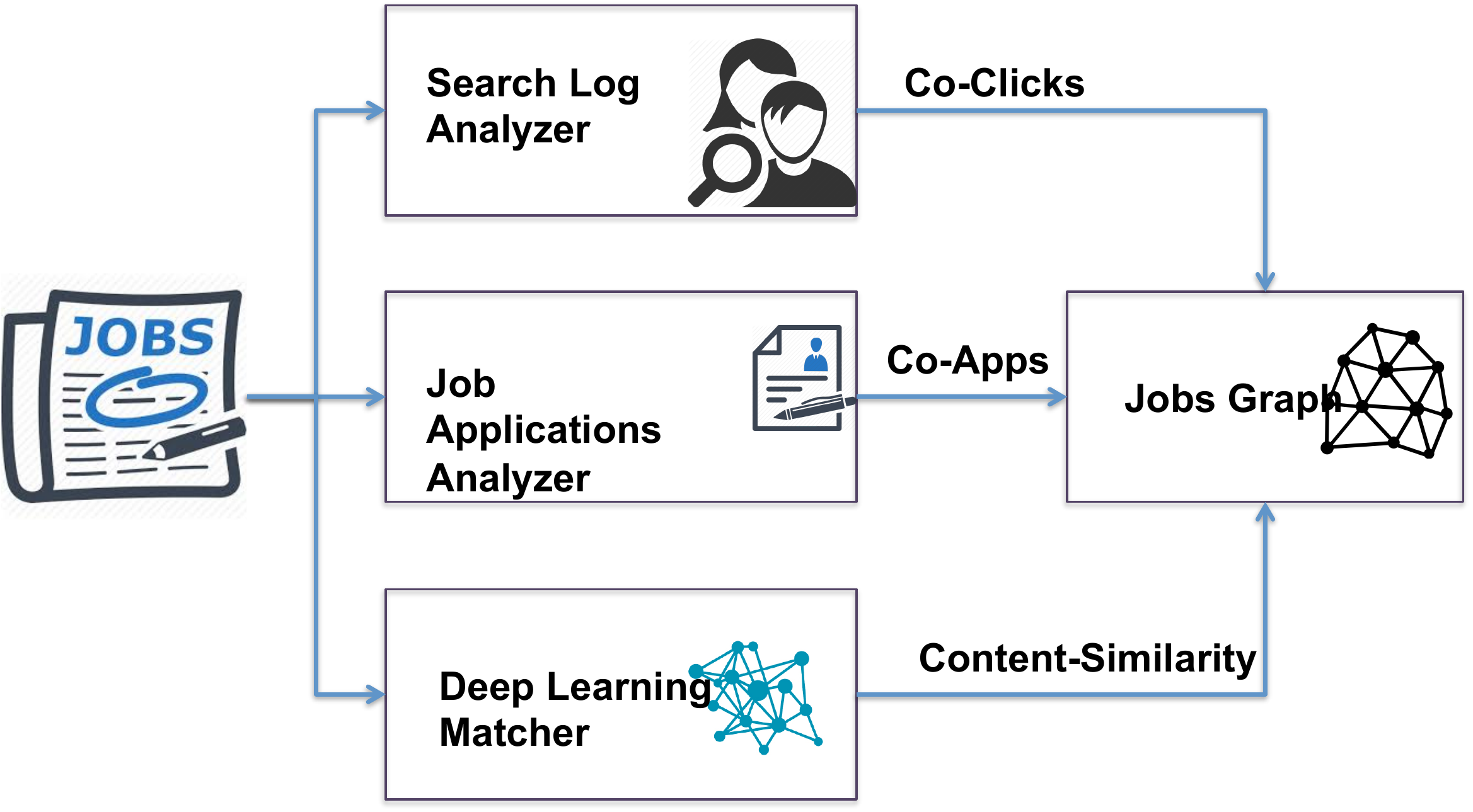,scale=0.36}\caption{\label{datasource} Data sources used to connect jobs in the recommendation graph}
	\label{fig:sources}
\end{figure}

\begin{figure*}
	%\begin{tabular}{@{}c@{}|c@{}|c@{}|c@{}|}
		\begin{tabular}{c|c|c|c}
		\includegraphics [scale=0.32] {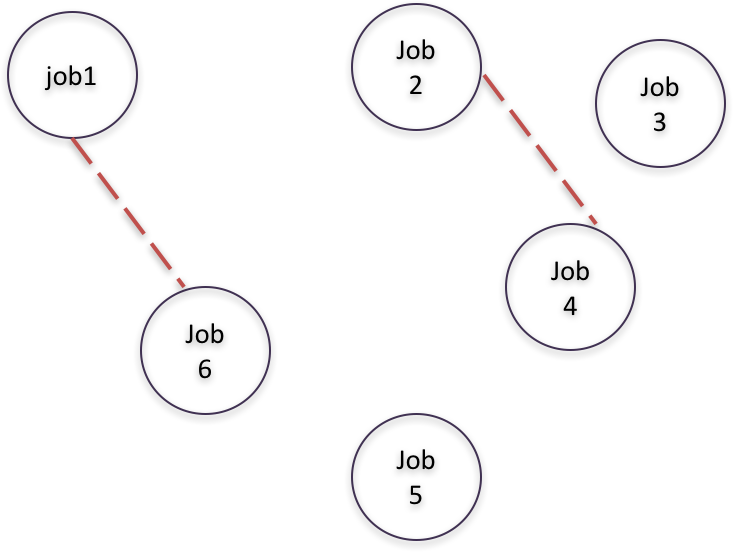}   & %\hspace{0.2cm} &
		%\hspace{0.2cm}
		\includegraphics [scale=0.32] {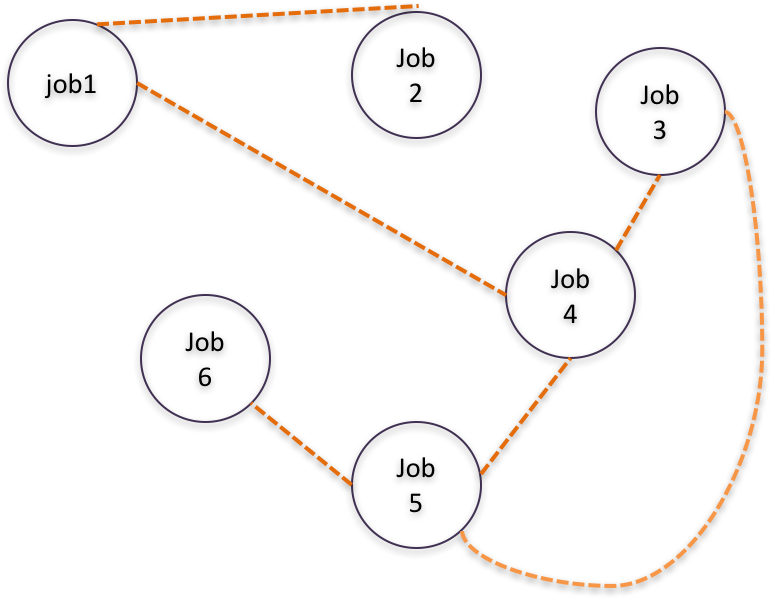}   &
		\includegraphics [scale=0.32] {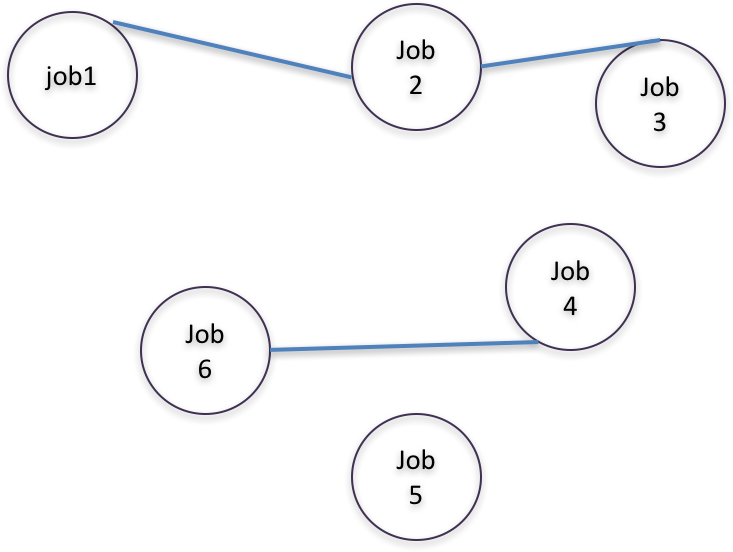}    & %\hspace{0.2cm} &
		%\hspace{0.3cm}
		
		\includegraphics [scale=0.32] {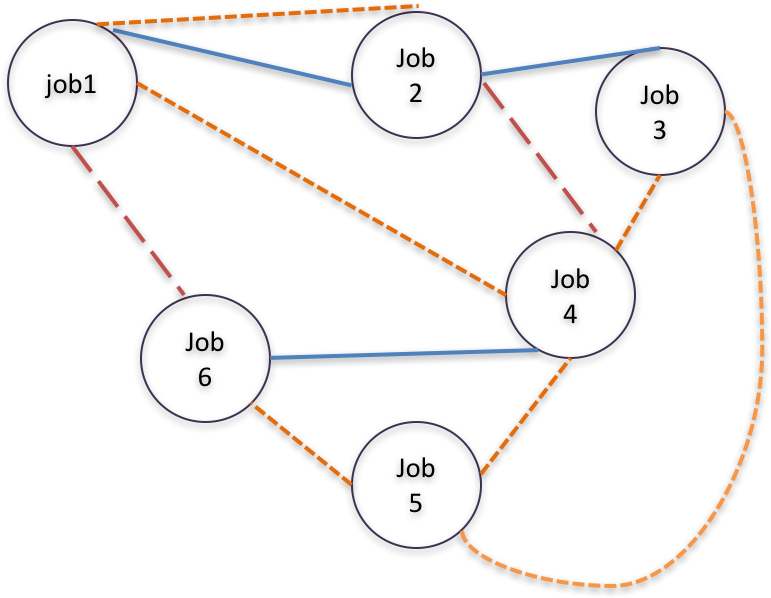}   

		\\
		(a) & (b)  & (c)  & (d)
	\end{tabular}
	\caption{ 
		The recommendation graph when connecting the nodes (jobs) using different sources of behavioral and contextual data. \textbf{(a)} using co-clicks, \textbf{(b)} using co-apps, and \textbf{(c)} using content-based similarity. Relying only on one source produces relatively sparse disconnected graph. Combining all behavioral and content-based signals in \textbf{(d)} produces denser graph.}
	
	\label{fig:edgs}
\end{figure*}

We propose a homogeneous Graph-Based Recommendation architecture (GBR). In our recommendation graph, jobs represent nodes and edges represent various similarity scores between pairs of jobs. We compute the similarity scores from multiple data sources that capture users behavior as well as resumes and jobs content (Figures \ref{fig:sources} \& \ref{fig:edgs}). We opted for building an item-based graph rather than a user-based or a user-item graph as it allows for more scalability given the number of users in our recommendation pool (hundreds of millions).

\subsection{Behavioral Data Sources}
We model users behavior from explicit signals such as their job applications, and implicit signals such as their clicks. Such user-job interactions allow us to compute different job-job co-statistics such as co-apps indicating, for any pair of jobs, how many users applied to both jobs, and co-clicks indicating, for any pair of jobs, how many users clicked both jobs when they appeared in the resultset of a user query.

%
%
%
%\begin{figure}
%		\centering \epsfig{file=Edges.pdf,scale=0.4}\caption{\label{edges} This figure depicts how the jobs recommendation graph look when connecting the nodes (jobs) using different source of behavioral and content data. In A the Co-Click signals only used to connect the nodes in the graph, however, most of the graph nodes are disconnected. Similar problem happen when relying on the other behavioral source of data (Co-Apps) in C. The content-based similarity in B generate a connected graph, but losing the powerful of the behavioral data. In D which represent our methodology, we build our graph by connecting the nodes with edges based on behavioral data as well as content similarity}
%	
%\end{figure}

First, we compute these co-statistics from a user-job bipartite multigraph where each edge is labeled with the type of user-job interaction (apply or click). After that, we construct our job-based labeled multigraph with jobs as nodes and multiple edges representing the co-apps and co-clicks counts. Formally, our co-statistics labeled multigraph is a 6-tuple graph $G=(V, E, \mathscr{T}_V, \mathscr{T}_E, \mathscr{L}_V, \mathscr{L}_E)$ such that:
\begin{itemize}
	\itemsep0em
	\item $V$ is a set of vertices representing jobs
	\item $E$ is a set of edges (arcs) connecting pairs of jobs.
	\item $\mathscr{T}_V$ is a finite set of tuples representing global statistics of each job (i.e. total applications and clicks).
	\item $\mathscr{T}_E$ is a finite set of tuples representing co-statistics of each edge connecting pairs of jobs (i.e. co-apps, and co-clicks counts).
	\item $\mathscr{L}_V: V \rightarrow \mathscr{T}_V$ and $\mathscr{L}_E: E \rightarrow \mathscr{T}_E$ are two maps describing the labeling of the vertices (jobs) and edges (co-statistics).
\end{itemize}

Our system utilizes several job-job association scores which could be computed directly from our initial job-based multigraph $G$.

\subsubsection{Maximum Likelihood Estimation (MLE)}
The conditional probability $p(i|j)$ of job $i$ given job $j$ can be viewed as a confidence score for recommending $i$ to a user who previously interacted with $j$. We estimate that conditional probability using the Maximum Likelihood Estimation (MLE) as in equation 1:
\begin{equation}
\hat{p}(i|j) = \frac{c(i,j)}{c(j)}
\end{equation}
where $c(i,j)$ is the co-statistic count between $i$ and $j$, and $c(j)$ is $j$'s statistic total count. As there are two types of co-statistics in $G$, we obtain, for each job pair, two scores $\hat{p_a}$ and $\hat{p_c}$ corresponding to co-apps and co-clicks respectively. As we notice, these scores are asymmetric (i.e. $\hat{p}(i|j)\ne\hat{p}(j|i)$), therefore, they will create directed edges when added to $G$ making it a multidigraph.

\subsubsection{Pointwise Mutual Information (PMI)}
%Normalized (pointwise) mutual information in collocation extraction
The conditional probability is biased by the popularity of the job to be recommended. Popular jobs which have been in the recommendation pool for several days will have more chance to be recommended than new jobs which received fewer interactions. To overcome this problem, we utilize Pointwise Mutual Information (PMI)~\cite{bouma2009normalized} as an additional correlation score. PMI normalizes the correlation score and provides a chance for the less popular jobs to appear at the top of the related jobs list if they are strongly related to the given job.  We compute for each job pair ($i,j$) a variant of PMI called PMI\textsuperscript{2} as in equation 2:
\begin{equation}
pmi^2(i,j) = log  \frac{c(i,j)^2}{c(i) \times c(j)}
\end{equation}
where $c(i,j)$ is the co-statistic count between $i$ and $j$, and $c(.)$ is the statistic total count. As with MLE, we obtain, for each job pair, two scores $pmi^2_a$ and $pmi^2_c$ corresponding to co-apps and co-clicks respectively. 

\subsection{Content-based Data Sources}
Relying on behavioral data solely results in lower quality recommendations when there is no user-job interaction data. This is akin to the 'cold-start' problem which becomes prominent when new jobs and/or new users are introduced to the recommendation pool. A related problem is data sparsity which appears when the interaction data exists but is scarce causing a very sparse graph structure. We alleviate these problems using content-based features extracted from both job descriptions and user resumes.

\subsubsection{Job Descriptions}
In order to reduce the sparsity in our recommendation graph, we densify the graph by creating an edge representing a content-based similarity score between pairs of jobs using their descriptions through a Deep Learning Matcher (DLM)\cite{yuan2016solving}. DLM works by training a neural network to generate a distributed representation of each job (a.k.a. embeddings) using its description. Typically a job description includes the job title, required and favorable skills, qualifications, experience, location, and employer information. After we obtain the embedding vectors $\mathbf{v_i},\mathbf{v_j}$ for jobs $i,j$ respectively, we compute the embedding-based cosine similarity score between them as in equation 3:
\begin{equation}
sim_e(i,j) = \frac{\mathbf{v_i}.\mathbf{v_j}}{||\mathbf{v_i}||\,||\mathbf{v_j}||}
\end{equation}
In addition, we cut off all similarities below threshold $\gamma$ which we tune empirically. Thus, the content-based similarity edge will be created only between pairs of nodes whose $sim_e\ge\gamma$. In this way, we guarantee an association between jobs even when they do not share any user interactions.

\subsubsection{User Resume}
In our model, we distinguish between two types of 'cold-start' users. Brand \emph{new} users and \emph{passive} users who do not have any interaction with current active jobs, but have a history with expired jobs. Our proposed method ranks the relevant active jobs for the users by taking into account their activity with the expired jobs in the past and/or their skills from their resumes.

For brand \emph{new} users, we build the user's preference vector by classifying their resume content into pre-defined fine-grained job categories ~\cite{javed2015carotene}. For \emph{passive} users, we build the user's preference vector as a mixture of: i) active jobs which are most similar to expired jobs in the user's interaction history using the job embeddings model, and ii) active jobs under job categories matching the user's resume.

We use jobs in the user's preference vector to extract a semi-personalized subgraph of the current graph. Then, we apply PageRank on this subgraph to further improve our recommendations by incorporating the general popularity of the jobs within the subgraph.

%Generating user's preference vector using his resume and/or past activity guarantees producing personalized recommendations for all users.

\subsection{Score Aggregation}
After computing all similarity scores from behavioral and content-based data sources. We create a single aggregated edge between pairs of nodes (jobs) representing the weighted aggregate correlation score using equation 4:
\begin{equation}
	corr(i,j) = w_1\!\sum_{s \in \{a,c\}}\!\hat{p_s}(i|j) + w_2\!\sum_{s \in \{a,c\}}\!pmi_s^2(i,j) + w_3 sim_e(i,j) 
\end{equation}

Where $w_1$, $w_2$, and $w_3$ are weights representing the importance of each similarity score and are tuned empirically. As we can notice, we aggregate scores from implicit and explicit behavioral signals as well as content-based signals. It is important to note that $corr(i,j)$ score is asymmetric and is computed in both directions resulting in two directed edges for each pair of jobs $i$ and $j$.

\subsection{Job Recommendation}

Since our proposed graph-based recommendation system is item-based, an effective way of recommending jobs to users is needed.  There are three types of users that are present in the system based on their behavior for the last 180 days:

1) Active users who have applied to or clicked on any job (currently active or expired) within the last 180 days.

2) New users and passive users who have done neither within the last 180 days, but either have an uploaded resume on their account at CareerBuilder.com, or otherwise specified the category of jobs they are looking for. % This is a good example of the `Cold-Start` problem for CF approaches.

3) New Users who have not yet interacted with any jobs on the site, uploaded a resume, or have specified their preferred category of jobs.

Since we also want to incorporate active and expired job listings in order to have as rich patterns as possible to aide the score aggregation process, and to reduce the sparsity of the multigraph, we add all jobs (active and expired) that were created within the last 180 days to the GBR with all their statistics and co-statistics. But in order to recommend active jobs only, in the score aggregation stage, we only generate directed edges whose destination is an active job. Restricting the content of the multigraph to data from the last 180 days ensures the scalability of our approach. Shorter (or longer) periods could be considered as long as they yield rich enough patterns to carry out proper recommendations.

For active users (type 1), generating recommendations is straightforward: given the jobs (active or expired) they have interacted with within the last 180 days (source jobs), recommend their highest similar jobs according to the aggregated score tuples generated by the GBR system. Since the user could have had interacted with multiple jobs (source jobs), and each of those jobs provides its own recommendations via the aggregated scores, a re-ranking step takes place based on how recent the interaction with this source job was; an activity score is introduced to each user's interaction with a source job: the more recent the interaction, the higher this activity score is. The aggregated scores of the recommended jobs coming from each source job are multiplied with the respective source job's activity score, hence re-ranking them in an intuitive manner. This allows for jobs similar to the user's recent interactions to be at the top, while still not dismissing recommendations based on older interactions.

\begin{figure}
	%\begin{tabular}{@{}c@{}|c@{}|c@{}|c@{}|}
	\begin{tabular}{c|c}
		\includegraphics [scale=0.32] {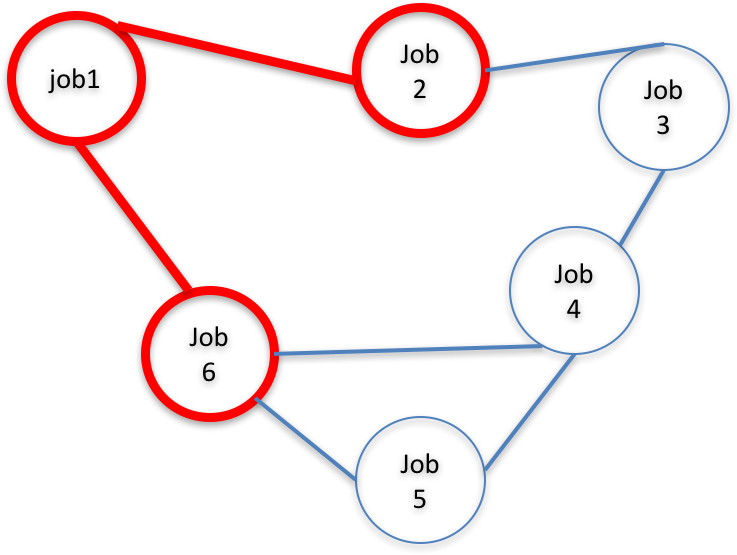}   & %\hspace{0.2cm} &
		%\hspace{0.2cm}
		
		\includegraphics [scale=0.32] {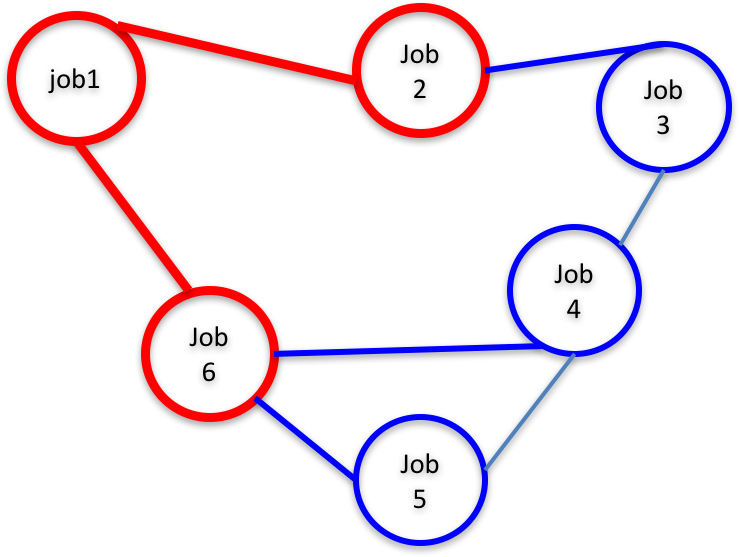}   
		
		\\
		(a) & (b)  
	\end{tabular}
	\caption{ 
		\textbf{(a)} the recommendations of job 1 are very few. No direct links between 1 and any of its neighbors' neighbors. \textbf{(b)} Extending the recommendations of job 1 to include jobs 3,4, and 5 through 2 and 6. The aggregate score tuples for non-adjacent jobs is set through multiplying the aggregate scores along the path. Eg., $score(1\!\rightarrow\!4)\!=\!score (1\!\rightarrow\!6) \times score(6\!\rightarrow\!4)$.}
	
	\label{fig:edgs_traverse}
\end{figure}

For some users, the aforementioned strategy does not yield enough recommended jobs, as their source jobs have only a few aggregate score tuples. As a solution, we go into a second iteration of recommending jobs from aggregate scores but this time, using the already just recommended jobs as the source jobs. Each time multiplying the aggregated scores along the path to preserve the priorities of the ranks. This process is illustrated by Figure \ref{fig:edgs_traverse}.

For passive and new users of type 2, the only source of clues we have about them is their processed resume and/or preferred job category. For those users, we take their fine-grained job category based on their resume, and through personalized PageRank, we get the list of active jobs in that job category ranked based on their popularity. From the resume information, a preferred or current location might be listed (state, city ..etc). We use a simple heuristic measure to re-rank the list of active jobs, favoring relevant jobs that are in locations near the location of the source job.

For users of type 3, we can only serve them global recommendations drawn from the global PageRank over the active jobs.

For some active users, if the 2nd level propagation still does not fill the minimum required number of jobs to recommend, then we use their resume information to generate more personalized PageRank recommendations, similar to what is done to passive users.

\begin{figure}
	\centering \epsfig{file=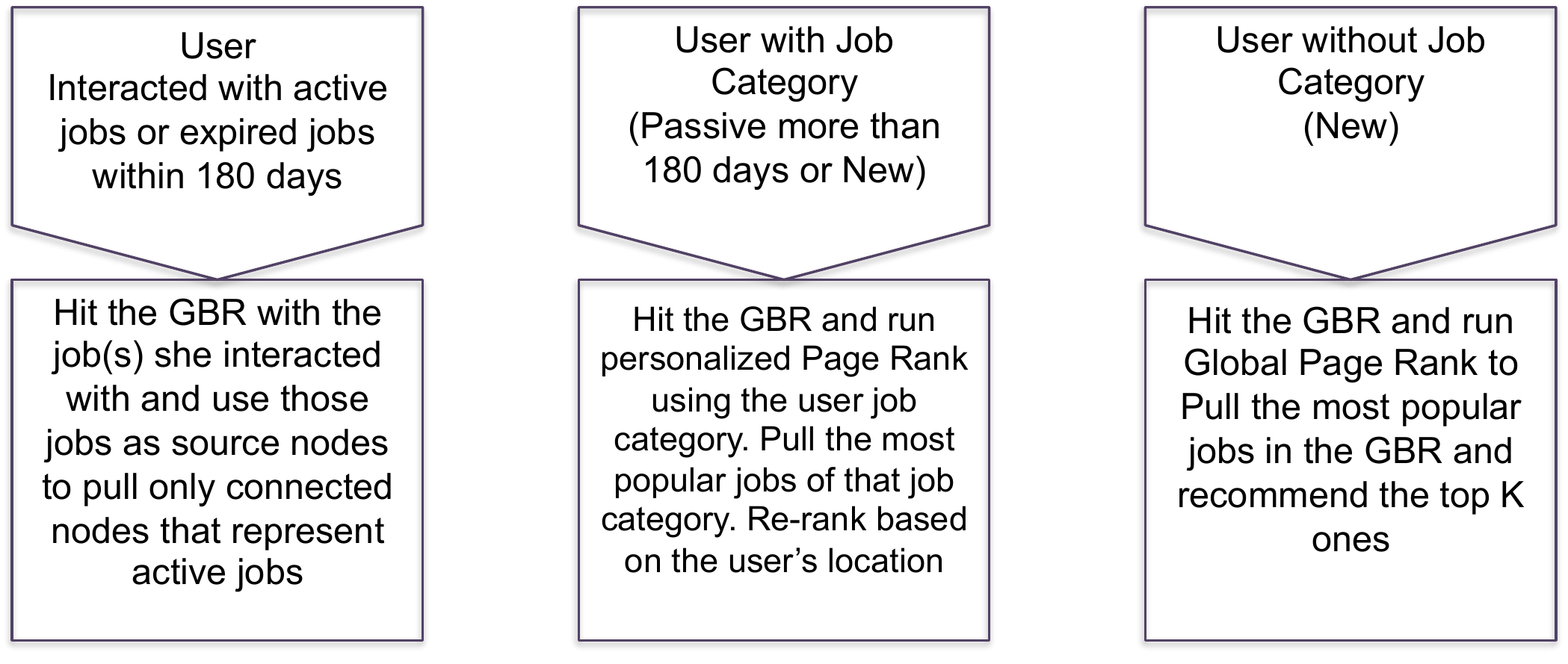,scale=0.40}\caption{\label{fig:scenarios} The 3 types of users and the process of generating their recommendations}
	
\end{figure}

Figure \ref{fig:scenarios} summarizes the process of generating recommendations for the three types of users.

\section{Experiments and Results  }

\begin{table}
	\centering

\caption{  Random Sample job-job Recommendations }
\label{manual-eval}
\begin{tabular}{p{4cm}|p{4cm}}
	\hline
	\hline
		\bfseries applied job  &  	\bfseries top recommended job \\
	\hline
	\hline
Staffing Consultant / Recruiter (Olde West Chester, OH) & 	Staffing Recruiter/Onsite (Cincinnati, OH) \\ 	\hline
Certified Nursing Assistant / CNA (Evansville, IN)	&	Certified Nursing Assistant (CNA) (Evansville, IN) \\ 	\hline
Staff Accountant (Ventura, CA)& 	Staff Accountant (Camarillo, CA) \\ 	\hline
Sales Representative (Cherry Hill, NJ)	&	Sales Representative (Oaklyn, NJ) \\ 	\hline
Freight Operations Supervisor (Fort Worth, TX) &	Import Supervisor (Dallas, TX)\\ 	\hline
Call Center Supervisor (Memphis, TN)	&	Customer Service Supervisor - Call Center (Memphis, TN) \\ 	\hline
Armed and Unarmed Security Guards (Hot Springs, AR) & Security Guards (Arkadelphia, AR) \\ 	\hline
Electrical Engineer (Saratoga Springs, NY) & Professional Electrical Engineer (Malta, NY) \\ 	\hline
Office Assistant (Lancaster, PA)	& Office Assistant (Harrisburg, PA) \\ 	\hline
Entry Level Child Care Giver \& Teacher (Live-in Social Services)	(Minneapolis, MN) & Entry Level Child Care giver Teacher (Minneapolis, MN) \\ 	\hline
	\hline
\end{tabular}
\end{table}

To test and evaluate our proposed system, we used a real dataset (jobs and users) via CareerBuilder.com. CareerBuilder operates the largest job posting board in the U.S. and has an extensive growing global presence, with millions of job postings, more than 60 million actively searchable resumes, over one billion searchable documents, and more than a million searches per hour. In this section, we show how we tested the effectiveness of our proposed graph-based job recommendation system as well as an A/B test against CareerBuilder's classical recommendation system.

\subsection{The GBR System Implementation}

We implemented our graph-based recommendation model using Apache Spark's GraphX framework.
The jobs data is stored in Hive databases which makes it convenient to interface with Spark. Through SparkSQL, we process the data and store the results in the shape of aggregated score tuples, without transferring the data around needlessly.

\subsection{Parameter Tuning}

The parameters of the aggregated scores equation, the DLM cutoff $\gamma$, the activity score re-ranking heuristic, and location-based re-ranking parameters of the recommended jobs were all tuned empirically through multiple rounds of trial and error. Experts and job recruiters at CareerBuilder were asked to validate rounds of recommendations for randomly sampled jobs obtained via different combination of these parameters. Combinations of entire signals were also considered to make sure no over-engineering was done, i.e., evaluating recommendations based on co-apps only, co-clicks only, DLM similarity only, as well as all their combinations.

\subsection{Qualitative Evaluation of GBR}

To show case the problem of sparseness (disconnectedness), we counted the active jobs that were connected to any other jobs through a co-statistic, hence appearing in at least one aggregated score tuple, and the results were as follows:
With co-clicks edges only, out of a sample of 500K jobs of the active jobs, only 28\% were connected to other jobs. With the co-apps edges alone, the number rises up to only around 40\%.  With both co-clicks and co-apps edges, the number barely reaches 61\% of active jobs. This truly shows the inherent issue of sparsity and the tough cold-start new jobs have. However, augmenting with the content-based similarity edges coming from the DLM, approximately 100\% of the active jobs are connected, only a handful are left out, even though we apply a cutoff threshold of $\gamma=0.4$ (i.e., no DLM edge constructed below 0.4). This allows for virtually no active job to be left out through a very meaningful signal.

We empirically evaluated samples of the aggregate score tuples of paired jobs resulting from augmenting the jobs' behavioural co-statistics with DLM similarities. We used jobs to which random users have recently applied, and validated the highest 50 jobs with which they got paired according to the GBR aggregate scores. For a typical job and 50 jobs paired with it: 43 would turn out to be very relevant matches, 2 passable recommendations (i.e., the pair would be in the same coarse-grained job category), and 5 bad matches (i.e., pairs that are unrelated). Counting the relevant and the passable matches together, on average, we get around 90\% empirical relevancy. Table \ref{manual-eval} shows a small batch of such pairs.

\subsection{User-Response Metrics}
In both of the upcoming testing sections, we focus on two main user response metrics. The first is the Expression of Interest (EOI) which is basically a count of how many new job applications resulted from the sent recommendation emails. For this, the recommended job link was augmented with tokens to track the source of the recommendation system that produced this recommended job. The second metric is the Click Through Rate (CTR) which is a count of how many recommended jobs' links got clicked, also augmented with a tracking token.  We are especially interested in EOI and CTR over the opened emails, as that would show how relevant were the recommended jobs to the users.

\subsection{Pilot tests}
We set up a recurring email campaigns with recommendations generated through the GBR system. The first campaign started with 10K users sampled from the available users in the system. Users were sent an email containing the top 15 recommended jobs they had not interacted with before, coming from the GBR based on other jobs with which they had previously interacted. In the case of passive users, and users who do not have enough job interactions, the recommendations were based on the fine-grained job category in which these users' resumes show they might be interested, coming from the personalized PageRank subsystem.

We collected users' responses to those emails over a period of few days, and then generated and sent the same users new recommendation emails taking into account all their possibly new interactions with the jobs in the system. We repeated the same process one more time, collecting the final metrics after another few days, the results of which appear in Table \ref{email-camp1}. 
%The number of users decrease each step as some users opt to unsubscribe to the email recommendation service.

With the very promising EOI and CTR responses of that first campaign, we executed another email campaign targeting 100K users this time. Table \ref{email-camp2} shows the user-response metrics of the 100K campaign which is still very promising especially in terms of EOI.

\begin{table}
	\centering
	\caption{GBR Email Campaigns (10k users) }
	\label{email-camp1}
	\begin{tabular}{c|c|c|c|c}
		\hline
		\hline
		\bfseries Sent &	 \bfseries Opens & \bfseries EOI &	\bfseries EOI to Open	& \bfseries CTR  \\
		\hline

		\hline
		9,968 &	1,670 & 598 &	35.81\%	&25.09\% \\
		
		\hline
		9,800 & 	1,810 &242	 &13.37\%&	20.28\% \\
		\hline
		9,740 &	1,600	&	225 &	14.06\% &	21.13\% \\
		
		\hline
		\hline
	\end{tabular}
	
\end{table}

\begin{table}
	\centering
	\caption{GBR Email Campaigns (100k users)}
	\label{email-camp2}
	\begin{tabular}{c|c|c|c|c}
		\hline
		\hline
		\bfseries Sent &	 \bfseries Opens & \bfseries EOI &	\bfseries EOI to Open	& \bfseries CTR \\
		\hline
		\hline
		99,320	 & 32,580	&	9,073	&27.85\%&	17.25\% \\	
		\hline
		81,000	 & 27,740 & 7,773	&28.02\% &	20.48\%	\\
		\hline
		71,230	 & 25,500 &		7,127&	27.95\%	 &18.31\% \\

		\hline
		\hline
	\end{tabular}
	
\end{table}

\subsection{A/B Test - GBR vs. CF}
\subsubsection{Experiment Setup}

To test the effectiveness of our proposed GBR, we carried out multiple A/B tests against the classical recommendation system CareerBuilder had been using at the time.   
The classical system is a custom CF model that is only based on co-apps. The CF system would track, for every job, the last 50 applicants. Whenever a user applies to a certain job, it would recommend jobs that, those other 50 applicants associated with this job, had applied to, recently.

The A/B test was as follows: the classical system would send recommendation emails to 1 million users sampled from the available users in the system, while the GBR would send recommendation emails to 350K different random users. To further rule out any biases that might tip the comparison off, both of the email campaigns were set to begin sending on the same day and at the same time period in order to not have any difference in the users' general email checking patterns, or any other external factors that might have any influence. All the user response metrics collection was cut off after 24 hours from the sending time.

\begin{table}
	\centering
	\caption{Email  Campaign results (GBR vs. CF) }
	\label{email-camp3}
	\begin{tabular}{c|c|c}
		\hline
		\hline
		\bfseries Metric & \bfseries GBR &\bfseries CF\\
		\hline
		\hline
		Sent & ~350K & ~1M\\
		\hline
		Open Rate & 22\% & 13\%\\
		\hline
		CTR & 14.7\% &15.3\%\\
		\hline
		EOI & \bfseries 18,023 & 15,439\\
		\hline
		EOI/Open &\bfseries 23.4\% &11\% \\
		\hline
		\hline
	\end{tabular}
\end{table}

\subsubsection{Populations}

The type of users sampled for the two email campaigns included seekers of a rather diverse types of jobs; ranging from health care, technology, office workers, machine operators, sales, and a lot more. We did not restrict the types of jobs as we wanted to compare the potential of both systems and their effectiveness of attending to all different kinds of jobs, and how potentially different job seeking habits among those job seekers might affect the relevance of the recommendations of one system or the other.

\subsubsection{Results}
The GBR surpassed the classical CF system by quite a margin as Table \ref{email-camp3} shows in terms of EOI/Open. Despite the number of GBR emails sent being almost only a third of what the CF system sent, GBR achieved 12\% more EOI/Open.  In addition, by looking at the absolute EOI metric, Table \ref{email-camp3} shows that we could achieve even larger number of EOI with a much smaller number of sent emails.

\subsection{Other Considered Baselines - Matrix Factorization}

We have also tried other baselines, specifically Matrix Factorization schemes.
\subsubsection{Premise and Mechanism}
 In such schemes, the users' interactions with the items in the system (jobs) are represented as a large-scale sparse matrix $R$ of size $(m,n)$, whose rows represent the $m$ users and columns represent the $n$ items. The  matrix entry $R_{i,j}$ represents a signal of how interested user $i$ in job $j$ on a particular scale (e.g., [0-1]). 
 
 Matrix factorization then assumes that the matrix A can be factorized into two matrices $U$ and $J$ of sizes $(m,k)$ and $(n,k)$ respectively, such that
 \begin{equation}
 R \approx U J^T
 \label{mat-fact}
 \end{equation}
 
In elaborate schemes \cite{matfactpaper}, additive biases are added to the dot product to relieve the representation vectors from catering to the various tendencies in the datasets. These biases are usually $\mu$ for a general signal mean, $b_u$ a bias for the user $u$ usual level of rating, $b_j$ a bias for item $j$ usual received rating. Therefore the dot product purpose now is to provide the deviation from the user's, the item's, and the overall tendencies. Hence the reconstructed signal would look like:
 \begin{equation}
r_{u,i} = (\mu + b_u + b_j + \mathbf{j}.\mathbf{u})
\label{mat-fact-svd}
\end{equation}

In even more elaborate schemes, implicit signals of the user's preference are also factored in, to account for the very few explicit signals new users provide.
\begin{equation}
r_{u,i} = (\mu + b_u + b_j + \mathbf{j}.\big[\mathbf{u} + |N(u)|^{-0.5}\sum_{j \in N(u)} \mathbf{y_j}\big])
\label{mat-fact-svd++}
\end{equation}
Where $N(u)$ are the items user $u$ had implicitly interacted with, and $\mathbf{y_j}$ is a vector to capture this interaction.
 The equivalence in equations \ref{mat-fact}, \ref{mat-fact-svd} and \ref{mat-fact-svd++} is only limited to the non-zero items in matrix A, or more specifically, in items where the original signals are known. 
 The resulting representative matrices U and J are collections of vectors in a subspace of $k$ dimensions (usually k is much smaller than M and N), representing the users and the jobs respectively. According to the premise depicted in equation \ref{mat-fact}, a dot product between user $i$ and job $j$ vectors should reveal $i$'s interest in $j$, or in other words, $j$'s relevance to $i$. The same for the elaborate schemes after adding up all the biases involved.

\subsubsection{producing recommendations}
A job recommendation system's objective is to find jobs a user might be interested in. Hence, after the factorization, 
the process of recommending jobs to a certain user boils down to finding the jobs whose vectors produce the highest relevance rate upon matrix-multiplied by the user's vector. And then filtering are reranking based on other criteria (whether the users has had interacted with those jobs before, location sensitivity...etc).

\subsubsection{Our Data and Setup}
 
 We used Apache Spark's MLLib implementation of ALS (Alternating Least Squares)~\cite{koren2009matrix}, to factorize an interaction matrix between 1.5M users and the jobs they interacted with within the last 6 months (160K jobs).
 
 We put a signal of 1.0 in cell $i,j$ of the interaction matrix, if user $i$ has had applied to job $j$. This by itself would render all reconstructed entries as 1.0, since now an entry is either unknown or 1.0, and a matrix of all 1.0s is the lowest-rank possible. To mitigate this mode collapse, a negative rating of -1.0 was used in cell $i,j$ if user $i$ has opened an email we had sent of jobs we recommended to her, but she did not click on job $j$ in that email.
 
 \subsubsection{Problems with this approach}
 Although we achieved a relatively low reconstruction MSE (Mean-Squared-Error). It was apparent that this approach is not suited for our needs for the following reasons:
 
 \paragraph{Cold-Start} The biggest problem with this approach is the cold-start problem. When a user has no interaction data within our system it is difficult to serve them recommendations. When no  users have interacted with a job, it is not possible to recommend it in any scenario. 
 This situation manifests in two ways: 
\begin{enumerate}
	\item Although the jobs in our system that have existed in the last 6 months far exceed 160K, the interaction of those 1.5M users only touched those 160K, and since all the signals in matrix factorization are based on interaction, jobs outside those 160K do not get to be represented by other than the trivial vector of zeros. 
	\item New users and new jobs that have no interactions yet are also out of luck in this recommendation scheme. And even when they get some interaction, they will have to wait for the next run of this large-scale intensive process to get represented.
\end{enumerate}
 
 Since GBR is a hybrid system, new jobs do not need to wait to be represented. Based on the job description, the DLM module learns its embedding vector which can be readily used to find similar jobs content-wise. Additionally, new users do not need to wait for the factorization process to take place in order to get profiled. The moment a user makes an interaction with a job (click or application), or just upload resume that gets classified, she can immediately get recommendations served and personalized through GBR's use cases.
 
 Although, like the GBR, the Asymmetric SVD model in \cite{matfactpaper} does not explicitly parametrize the users with their own vectors, but rather profiles them through the items they have interacted with, it still requires retraining to parametrize new items.

 \paragraph{Distribution Skew} An unavoidable characteristic of the users and jobs in our system is the non-uniform and highly-skewed distribution across job-types and users' interest in them. Since matrix factorization schemes seek to minimize the square error between the reconstructed signals and the original ones via low-rank representations, the optimizer gets forced to disregard relatively rare interactions (treat as noise) if the model is not flexible enough (e.g., $k$ not large enough). To mitigate this behavior, individual weights are assigned to each entry in the interaction matrix. These weights help penalize the reconstruction loss differently. If rare interactions (distribution-wise) get assigned higher weights than the rest, their reconstruction loss would contribute much more to the overall reconstruction loss of the interaction matrix. Thus forcing the optimizer to treat them more carefully. However, without adequate background of the user, it becomes a much harder problem to differentiate between legitimate rare signals and noise in order not to give noise higher weights as well.

 All of which makes matrix-factorization a computationally expensive process for a very dynamic system. These problems prevented us from going forward with an A/B test against the GBR as the recommendation pool of jobs here is very little and rare job types are poorly represented.

\section{Conclusion and Future Work}
Recommender system technologies have gained considerable popularity in online domains due to its effectiveness in creating commercial and social value. As one of such domains, online recruiting services utilize recommender systems to serve millions of applicants with relevant and personalized job postings. To bring the full potential of advanced recommender systems into the job search domain, we proposed novel methods for a scalable and robust job recommendation system. Our approach entails a multigraph of jobs connected by similarity edges which are tailored to capture their full relationship based on users behavior as well as jobs content. Moreover, we fully exploit the abundance of available implicit and explicit online data .
We empirically evaluate the recommendations coming out of the proposed approach and find that it achieves around 90\% accuracy, on average. We also show that our proposed system achieves higher EOI with a wide margin over the classical CF approach while needing only third of the number of sent emails.

%\subsection{Future work}
We are working to improve this approach even further to address three main issues:
\begin{itemize}

\item Many of the parameters in the system were decided heuristically and with manual evaluation.  For a dynamic system, such parameters are better learned in a way to maximize user's CTR (Click-Through-Rate). A Learning-To-Rank module could be utilized to learn to better rerank recommendations based on user's interactions with previously served recommendations. 

\item CareerBuilder serves job seekers in more than 24 countries with different spoken languages. Extending the content-based deep learning matcher to languages other than english in order to effectively bring the GBR to serve non-English speaking countries is a priority.  Deep learning approaches pave the way towards language agnostic NLP tools, so we will be looking to train more models to capture the similarity between job postings for different languages.

\item Although the sparseness of the behavioral graph is reduced by introducing the content-based similarity signal, it still results in jobs only connected through content-based similarity. A possible extension is to map the nodes of the graph (the jobs) to an embedding space that captures the behavioral pattern as well, since the context of the nodes is well-defined (in our case, it is the co-apps and the co-clicks), and then use this embedding to calculate the similarity between any two nodes to enrich the graph even further. This is a straightforward extension of the word2vec approach, and has been studied by others, notably \cite{MaMS16,GroverL16,VasileSC16}.

\end{itemize}

\section*{Acknowledgment}

The authors would like to deeply thank David Lin from CareerBuilder for providing very professional evaluation to the quality of the recommendations and working on analyzing the A/B test results. We also would like to thank the consumer email team at CareerBuilder for their help and support to run the large scale A/B test on hundred of thousands of recommendation emails.

\bibliographystyle{ieeetr}
\bibliography{gbr}

\begin{thebibliography}{10}

\bibitem{herlocker2004evaluating}
J.~L. Herlocker, J.~A. Konstan, L.~G. Terveen, and J.~T. Riedl, ``Evaluating
  collaborative filtering recommender systems,'' {\em ACM Transactions on
  Information Systems (TOIS)}, vol.~22, no.~1, pp.~5--53, 2004.

\bibitem{deshpande2004item}
M.~Deshpande and G.~Karypis, ``Item-based top-n recommendation algorithms,''
  {\em ACM Transactions on Information Systems (TOIS)}, vol.~22, no.~1,
  pp.~143--177, 2004.

\bibitem{sarwar2001item}
B.~Sarwar, G.~Karypis, J.~Konstan, and J.~Riedl, ``Item-based collaborative
  filtering recommendation algorithms,'' in {\em Proceedings of the 10th
  international conference on World Wide Web}, pp.~285--295, ACM, 2001.

\bibitem{adomavicius2005toward}
G.~Adomavicius and A.~Tuzhilin, ``Toward the next generation of recommender
  systems: A survey of the state-of-the-art and possible extensions,'' {\em
  IEEE transactions on knowledge and data engineering}, vol.~17, no.~6,
  pp.~734--749, 2005.

\bibitem{li2009grocery}
M.~Li, B.~M. Dias, I.~Jarman, W.~El-Deredy, and P.~J. Lisboa, ``Grocery
  shopping recommendations based on basket-sensitive random walk,'' in {\em
  Proceedings of the 15th ACM SIGKD}, pp.~1215--1224, ACM, 2009.

\bibitem{linden2003amazon}
G.~Linden, B.~Smith, and J.~York, ``Amazon. com recommendations: Item-to-item
  collaborative filtering,'' {\em IEEE Internet computing}, vol.~7, no.~1,
  pp.~76--80, 2003.

\bibitem{das2007google}
A.~S. Das, M.~Datar, A.~Garg, and S.~Rajaram, ``Google news personalization:
  scalable online collaborative filtering,'' in {\em Proceedings of the 16th
  international conference on World Wide Web}, pp.~271--280, ACM, 2007.

\bibitem{baluja2008video}
S.~Baluja, R.~Seth, D.~Sivakumar, Y.~Jing, J.~Yagnik, S.~Kumar,
  D.~Ravichandran, and M.~Aly, ``Video suggestion and discovery for youtube:
  taking random walks through the view graph,'' in {\em Proceedings of the 17th
  international conference on World Wide Web}, pp.~895--904, ACM, 2008.

\bibitem{yildirim2008random}
H.~Yildirim and M.~S. Krishnamoorthy, ``A random walk method for alleviating
  the sparsity problem in collaborative filtering,'' in {\em Proceedings of the
  2008 ACM conference on Recommender systems}, pp.~131--138, ACM, 2008.

\bibitem{yuan2016solving}
J.~Yuan, W.~Shalaby, M.~Korayem, D.~Lin, K.~AlJadda, and J.~Luo, ``Solving
  cold-start problem in large-scale recommendation engines: A deep learning
  approach,'' in {\em Proceedings of the 2016 IEEE International Conference on
  Big Data (Big Data)}, 2016.

\bibitem{breese1998empirical}
J.~S. Breese, D.~Heckerman, and C.~Kadie, ``Empirical analysis of predictive
  algorithms for collaborative filtering,'' in {\em Proceedings of the
  Fourteenth conference on Uncertainty in artificial intelligence}, pp.~43--52,
  Morgan Kaufmann Publishers Inc., 1998.

\bibitem{o1999clustering}
M.~O'Connor and J.~Herlocker, ``Clustering items for collaborative filtering,''
  in {\em Proceedings of the ACM SIGIR workshop on recommender systems},
  vol.~128, UC Berkeley, 1999.

\bibitem{shepitsen2008personalized}
A.~Shepitsen, J.~Gemmell, B.~Mobasher, and R.~Burke, ``Personalized
  recommendation in social tagging systems using hierarchical clustering,'' in
  {\em Proceedings of the 2008 ACM conference on Recommender systems},
  pp.~259--266, ACM, 2008.

\bibitem{koren2009matrix}
Y.~Koren, R.~Bell, and C.~Volinsky, ``Matrix factorization techniques for
  recommender systems,'' {\em Computer}, vol.~42, no.~8, 2009.

\bibitem{candillier2007comparing}
L.~Candillier, F.~Meyer, and M.~Boull{\'e}, ``Comparing state-of-the-art
  collaborative filtering systems,'' in {\em International Workshop on Machine
  Learning and Data Mining in Pattern Recognition}, pp.~548--562, Springer,
  2007.

\bibitem{lops2011content}
P.~Lops, M.~De~Gemmis, and G.~Semeraro, ``Content-based recommender systems:
  State of the art and trends,'' in {\em Recommender systems handbook},
  pp.~73--105, Springer, 2011.

\bibitem{chen2005link}
H.~Chen, X.~Li, and Z.~Huang, ``Link prediction approach to collaborative
  filtering,'' in {\em Digital Libraries, 2005. JCDL'05. Proceedings of the 5th
  ACM/IEEE-CS Joint Conference on}, pp.~141--142, IEEE, 2005.

\bibitem{page1999pagerank}
L.~Page, S.~Brin, R.~Motwani, and T.~Winograd, ``The pagerank citation ranking:
  Bringing order to the web.,'' tech. rep., Stanford InfoLab, 1999.

\bibitem{zhou2007bipartite}
T.~Zhou, J.~Ren, M.~Medo, and Y.-C. Zhang, ``Bipartite network projection and
  personal recommendation,'' {\em Physical Review E}, vol.~76, no.~4,
  p.~046115, 2007.

\bibitem{aggarwal1999horting}
C.~C. Aggarwal, J.~L. Wolf, K.-L. Wu, and P.~S. Yu, ``Horting hatches an egg: A
  new graph-theoretic approach to collaborative filtering,'' in {\em
  Proceedings of the fifth ACM SIGKDD international conference on Knowledge
  discovery and data mining}, pp.~201--212, ACM, 1999.

\bibitem{pan2004automatic}
J.-Y. Pan, H.-J. Yang, C.~Faloutsos, and P.~Duygulu, ``Automatic multimedia
  cross-modal correlation discovery,'' in {\em Proceedings of the tenth ACM
  SIGKDD international conference on Knowledge discovery and data mining},
  pp.~653--658, ACM, 2004.

\bibitem{gori2006research}
M.~Gori and A.~Pucci, ``Research paper recommender systems: A random-walk based
  approach,'' in {\em Web Intelligence, 2006. WI 2006. IEEE/WIC/ACM
  International Conference on}, pp.~778--781, IEEE, 2006.

\bibitem{haveliwala2003analytical}
T.~Haveliwala, S.~Kamvar, and G.~Jeh, ``An analytical comparison of approaches
  to personalizing pagerank,'' tech. rep., Stanford, 2003.

\bibitem{kim2011personalized}
H.-N. Kim and A.~El~Saddik, ``Personalized pagerank vectors for tag
  recommendations: inside folkrank,'' in {\em Proceedings of the fifth ACM
  conference on Recommender systems}, pp.~45--52, ACM, 2011.

\bibitem{haveliwala2002topic}
T.~H. Haveliwala, ``Topic-sensitive pagerank,'' in {\em Proceedings of the 11th
  international conference on World Wide Web}, pp.~517--526, ACM, 2002.

\bibitem{zhao2015skill}
M.~Zhao, F.~Javed, F.~Jacob, and M.~McNair, ``Skill: A system for skill
  identification and normalization.,'' in {\em AAAI}, pp.~4012--4018, 2015.

\bibitem{singh2010prospect}
A.~Singh, C.~Rose, K.~Visweswariah, V.~Chenthamarakshan, and N.~Kambhatla,
  ``Prospect: a system for screening candidates for recruitment,'' in {\em
  Proceedings of the 19th ACM international conference on Information and
  knowledge management}, pp.~659--668, ACM, 2010.

\bibitem{guo2016resumatcher}
S.~Guo, F.~Alamudun, and T.~Hammond, ``R{\'e}sumatcher: A personalized
  r{\'e}sum{\'e}-job matching system,'' {\em Expert Systems with Applications},
  vol.~60, pp.~169--182, 2016.

\bibitem{lee2007architecture}
I.~Lee, ``An architecture for a next-generation holistic e-recruiting system,''
  {\em Communications of the ACM}, vol.~50, no.~7, pp.~81--85, 2007.

\bibitem{farber2003automated}
F.~F{\"a}rber, T.~Weitzel, and T.~Keim, ``An automated recommendation approach
  to selection in personnel recruitment,'' {\em AMCIS 2003 proceedings},
  p.~302, 2003.

\bibitem{diaby2013toward}
M.~Diaby, E.~Viennet, and T.~Launay, ``Toward the next generation of
  recruitment tools: an online social network-based job recommender system,''
  in {\em Proceedings of the 2013 IEEE/ACM International Conference on Advances
  in Social Networks Analysis and Mining}, pp.~821--828, ACM, 2013.

\bibitem{rafter2000automated}
R.~Rafter, K.~Bradley, and B.~Smyth, ``Automated collaborative filtering
  applications for online recruitment services,'' in {\em International
  Conference on Adaptive Hypermedia and Adaptive Web-Based Systems},
  pp.~363--368, Springer, 2000.

\bibitem{lu2013recommender}
Y.~Lu, S.~El~Helou, and D.~Gillet, ``A recommender system for job seeking and
  recruiting website,'' in {\em Proceedings of the 22nd International
  Conference on World Wide Web}, pp.~963--966, ACM, 2013.

\bibitem{malinowski2006matching}
J.~Malinowski, T.~Keim, O.~Wendt, and T.~Weitzel, ``Matching people and jobs: A
  bilateral recommendation approach,'' in {\em System Sciences, 2006. HICSS'06.
  Proceedings of the 39th Annual Hawaii International Conference on}, vol.~6,
  pp.~137c--137c, IEEE, 2006.

\bibitem{paparrizos2011machine}
I.~Paparrizos, B.~B. Cambazoglu, and A.~Gionis, ``Machine learned job
  recommendation,'' in {\em Proceedings of the fifth ACM Conference on
  Recommender Systems}, pp.~325--328, ACM, 2011.

\bibitem{bouma2009normalized}
G.~Bouma, ``Normalized (pointwise) mutual information in collocation
  extraction,'' in {\em Proceedings of the Biennial GSCL Conference}, vol.~156,
  2009.

\bibitem{javed2015carotene}
F.~Javed, Q.~Luo, M.~McNair, F.~Jacob, M.~Zhao, and T.~S. Kang, ``Carotene: A
  job title classification system for the online recruitment domain,'' in {\em
  Big Data Computing Service and Applications (BigDataService), 2015 IEEE First
  International Conference on}, pp.~286--293, IEEE, 2015.

\bibitem{matfactpaper}
Y.~Koren, ``Factorization meets the neighborhood: a multifaceted collaborative
  filtering model,'' in {\em Proceedings of the 14th {ACM} {SIGKDD}
  international conference on Knowledge discovery and data mining},
  pp.~426--434, ACM, 2008.

\bibitem{MaMS16}
Q.~Ma, S.~Muthukrishnan, and W.~Simpson, ``App2vec: Vector modeling of mobile
  apps and applications,'' in {\em 2016 {IEEE/ACM} International Conference on
  Advances in Social Networks Analysis and Mining, {ASONAM} 2016, San
  Francisco, CA, USA, August 18-21, 2016}, pp.~599--606, ACM, 2016.

\bibitem{GroverL16}
A.~Grover and J.~Leskovec, ``node2vec: Scalable feature learning for
  networks,'' in {\em Proceedings of the 22nd {ACM} {SIGKDD} international
  conference on Knowledge discovery and data mining}, pp.~855--864, ACM, 2016.

\bibitem{VasileSC16}
F.~Vasile, E.~Smirnova, and A.~Conneau, ``Meta-prod2vec: Product embeddings
  using side-information for recommendation,'' in {\em Proceedings of the 10th
  {ACM} Conference on Recommender Systems, Boston,MA, USA, September 15-19,
  2016}, pp.~225--232, ACM, 2016.

\end{thebibliography}
 % sigproc.bib is the name of the Bibliography in this case% You must have a proper ".bib" file%  and remember to run:% latex bibtex latex latex% to resolve all references% ACM needs 'a single self-contained file'!%APPENDICES are optional%\balancecolumns

\end{document}